\documentclass[prl,showpacs,amsmath,amssymb,twocolumn,superscriptaddress,A4]{revtex4}

\usepackage{amssymb}
\usepackage{graphicx}
\usepackage{color}
\usepackage{dcolumn}
\usepackage{bm}
\usepackage{multirow}
\makeatletter
\def\NAT@spacechar{\,}
\makeatother

\begin{document}

\newcommand{\kfa}{KFe$_2$As$_2$}
\newcommand{\bkfa}{Ba$_{1-x}$K$_x$Fe$_2$As$_2$}
\newcommand{\tc}{$T_c$}
\newcommand{\hc}{$H_{c2}(T)$}
\newcommand{\ttt}{$T_0$}

\title{Pauli-Limited Multiband Superconductivity in \kfa}

\author{D. A. Zocco}
\email[E-mail: ]{diego.zocco@kit.edu}
\author{K. Grube}
\email[E-mail: ]{kai.grube@kit.edu}
\author{F. Eilers}
\author{T. Wolf}
\affiliation{Institute of Solid State Physics (IFP), Karlsruhe Institute of Technology, D-76021 Karlsruhe, Germany.}
\author{H. v.\,L\"{o}hneysen}
\affiliation{Institute of Solid State Physics (IFP), Karlsruhe Institute of Technology, D-76021 Karlsruhe, Germany.}
\affiliation{Physikalisches Institut, Karlsruhe Institute of Technology, D-76031 Karlsruhe, Germany}

\begin{abstract}
The upper critical field \hc\ of the multiband superconductor \kfa\ has been studied via low-temperature thermal expansion and magnetostriction measurements. We present compelling evidence for Pauli-limiting effects dominating \hc\ for $H$\,$\parallel$\,$a$, as revealed by a crossover from second- to first-order phase transitions to the superconducting state in the magnetostriction measurements down to 50\,mK. Corresponding features were absent for $H$\,$\parallel$\,$c$. To our knowledge, this crossover constitutes the first confirmation of Pauli limiting of the \hc\ of a multiband superconductor. The results are supported by modeling Pauli limits for single-band and multiband cases.
\end{abstract}


\maketitle

The upper critical field curve \hc\ of a type-II superconductor (SC) reflects basic properties such as pair-breaking mechanisms, Fermi surface (FS) anisotropies, and multiband effects. Spin-singlet superconductivity can be suppressed with magnetic fields by either forcing the charge-carrier motion into cyclotron orbits or by spin polarization of the quasiparticles (Zeeman splitting) \cite{matsuda07a}. \hc\ is usually limited by the first effect, commonly referred to as orbital pair-breaking, with the limiting field $\mu_0 H_{c2}^{orb}$\,=\,$\Phi_0/2\pi\xi^2$ for $T$\,=\,0, where $\Phi_0$ is the flux quantum and $\xi$ is the coherence length at $T$\,$\rightarrow$\,$0$, and it is mainly controlled by the slope $H_{c2}^{'}$\,=\,$dH_{c2}/dT|_{T_c}$, inversely proportional to the Fermi velocity of the quasiparticles. The second effect, commonly called Pauli pair-breaking, is characterized by $H_{c2}^{\mathrm{P}}$, determined from equating the superconducting condensation energy with the magnetic energy $(1/2)\mu_0 \chi_N (H_{c2}^{\mathrm{P}})^2$ (Chandrasekhar–-Clogston limit), where $\chi_N$ is the normal-state spin susceptibility. Pauli-limiting effects become important when the orbital shielding currents are reduced due to low-dimensional electronic structures or when $\chi_N$ is enhanced due to spin-orbit coupling. In these cases, $H_{c2}^{\mathrm{P}}$ can be smaller than $H_{c2}^{orb}$, and if the Maki parameter defined as $\alpha_\mathrm{M}$\,=\,$\sqrt{2} H_{c2}^{orb}/H_{c2}^{\mathrm{P}}$ becomes larger than 1.85, superconductivity becomes Pauli limited with a discontinuous transition at high fields \cite{maki66a}. In this field region and for clean-limit superconductors, the Zeeman splitting of the FS is expected to lead to a spatially modulated superconducting state, the so-called FFLO phase, predicted nearly 50 years ago independently by Fulde and Ferrell \cite{FF64a}, and Larkin and Ovchinnikov \cite{LO64a}.

Up to now, only few SCs are known in which Pauli-limiting effects are strong enough to induce a change from a second-order (SO) to a first-order (FO) phase transition. Some examples include heavy-fermion and organic SCs \cite{bianchi03a,lortz07a}. The existence of an FFLO state in these systems remains, however, under debate \cite{zwicknagl10a}. These Pauli-limited SCs have been consistently described as single-band systems. A challenging issue is the possibility of strong Pauli-limiting effects in multiband superconductors \cite{fuchs09a}. In these materials, bands contributing to the FS might have different dimensionality, and thus the condition for a discontinuous phase transition or FFLO state might differ from band to band. Theoretical calculations have predicted that the high-field $H_{c2}(T)$ of multiband systems should show pronounced deviations from that of single-band SCs \cite{gurevich10a}.

The iron-based multiband SCs present a unique opportunity to study these matters in detail. Here, we present measurements on \kfa\ single crystals, which give evidence for a Pauli-limited multiband SC. \kfa\ crystallizes in a tetragonal ThCr$_2$Si$_2$-type structure (space group $I$4/$mmm$). It is the end-member of the \bkfa\ series, in which the superconducting state reaches a maximum \tc\ of 38\,K at $x$\,$\sim$\,0.4 \cite{rotter08a}. Due to the proximity of these compounds to antiferromagnetic order, their pairing mechanism is believed to arise from magnetic fluctuations, as it is discussed for cuprate and heavy-fermion SCs \cite{castellan11a}. For \kfa, evidence for multigap nodal $s$-wave superconductivity has indeed been found in nuclear quadrupole resonance \cite{fukazawa09a} and angle-resolved photoemission spectroscopy (ARPES) \cite{okazaki12a}, while recent experiments suggest $d$-wave pairing \cite{hashimoto10a,reid12a,tafti13a}.

Compared to optimally doped \bkfa, the low superconducting transition temperature \tc\,$\sim$\,3.4\,K of \kfa\ allows us to explore its entire $H$-$T$ phase diagram. We performed thermal-expansion and magnetostriction measurements in a temperature range between 50\,mK and 4\,K and in magnetic fields up to 14\,T applied parallel and perpendicular to the $c$ axis of the crystals. Our experiments constitute an extension of the measurements performed above 2\,K by Burger $et\,al.$ \cite{burger13a}, in which initial evidence of strong Pauli-limiting effects was presented. The experiments were carried out in a home-built capacitive dilatometer. The linear thermal-expansion and magnetostriction coefficients are defined as $\alpha_{i}$\,=\,$L_i^{-1}\partial L_i / \partial T$ and $\lambda_{i}$\,=\,$L_i^{-1}\partial L_i / \partial (\mu_0 H)$, respectively, where $L_i$ is the length of the sample along the $i$\,=\,$a,c$ axis. As $\alpha_{i}$ is related to the uniaxial pressure dependence of the entropy $S$ via Maxwell relations, we can use $\alpha_{i}$ to search for nearby pressure-induced instabilities. Single crystals of \kfa\ were grown in a K-Fe-As melt rich in K and As to reduce the amount of magnetic impurities \cite{sm}. The residual resistivity ratio (RRR) of the samples amounts to $\sim$\,1000 \cite{hardy13a}. As flux-grown iron arsenides tend to form foliated stacks with embedded flux, the observation of quantum oscillations (QOs) for both field directions in our magnetostriction measurements represents a particularly reliable quality probe. The mean-free-paths (mfp) determined from the Dingle temperatures of the QOs amount to $\ell_{ab}=(177\pm 8)\,$nm and  $\ell_{c}=(52\pm 3)\,$nm along the $a$- and $c$ axis, respectively. With the coherence lengths of $\xi_{ab}$\,$\sim$\,15\,nm and $\xi_c$\,$\sim$\,3\,nm \cite{terashima09a}, the ratio $\ell / \xi$\,$\sim$\,15 confirms that the samples are in the superconducting clean limit. The extracted high effective masses are consistent with the enhanced Sommerfeld coefficient. Furthermore, the FS cross-sectional areas inferred from our data are in agreement with the reported electronic structure in which the contribution of each band to the FS differs in its dimensionality \cite{terashima10a}. Further details about the QOs of the magnetostriction will be given in a separate publication.

\begin{figure}
\includegraphics[width=3in]{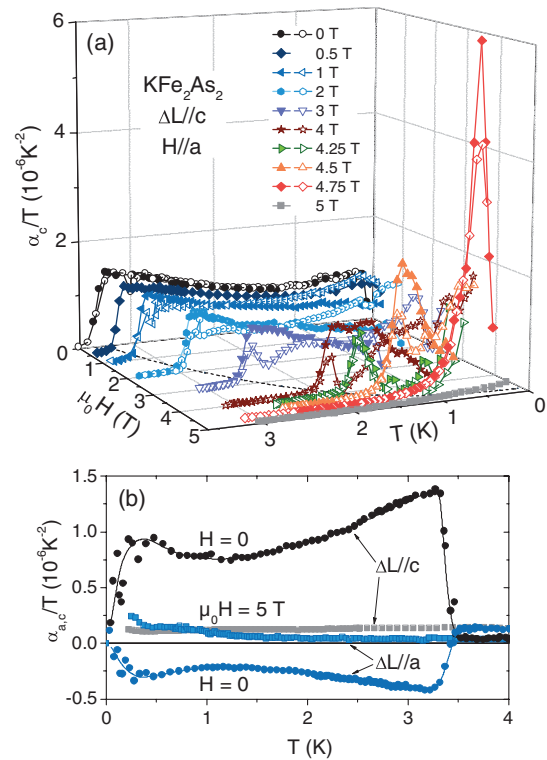}
\caption{(color online). (a) Thermal-expansion divided by temperature $\alpha_c /T$ at fields $H$\,$\parallel$\,$a$ ranging from 0\,T to 5\,T (full symbols: cooling, open symbols: heating). (b) $\alpha_c /T$ and $\alpha_a /T$ vs $T$ for $\mu_0 H$\,=\,0 and 5\,T of two different samples.}
\label{fig1}
\end{figure}

The linear thermal-expansion coefficients $\alpha_c/T$ of \kfa\ are plotted in Fig.\,\ref{fig1}(a) for $H$\,$\parallel$\,$a$. For $H$\,=\,0, the SC transition has a step-like form, with no appreciable difference between cooling and heating curves. Besides the steps at \tc, the data for $H$\,=\,$0$ show additional broad maxima at $\sim$\,0.5\,K, displayed in more detail in Fig.\,\ref{fig1}(b) for both $\alpha_a/T$ and $\alpha_c/T$. These features directly manifest the multiband nature of superconductivity in \kfa. A shoulder of $C/T$ vs $T$ has previously been observed in \kfa\ \cite{hardy13a}, similar to that of the well-known multiband SC MgB$_2$, in which the observed feature is caused by the opening of a low-energy superconducting gap on one of the weakly coupled bands \cite{bouquet01a}. Even though $\alpha_i/T$ and $C/T$ are interrelated via the Gr\"{u}neisen parameter, the maxima in $\alpha_i/T$ are much more pronounced than the ones observed in $C/T$.

For $H$\,$>$\,0 [Fig.\,\ref{fig1}(a)], the system enters an irreversible regime, possibly due to vortex pinning effects. As $H$ is increased and \tc\ is suppressed, a clear increase of $\alpha_c/T$ emerges at $\mu_0 H$\,=\,4\,T (\tc\,$\sim$\,1.7\,K) and continues to develop to a peak-like transition at higher fields. The increase of $\alpha(T,H)/T$ for large fields resembles a crossover from a SO to a FO phase transition, expected for a system presenting strong Pauli-limiting effects. Evidence for Pauli-limiting effects in \kfa\ has been reported in earlier measurements of \hc\ \cite{terashima09a, terashima13a} and magnetization \cite{burger13a}. For $H$\,$>$\,$H_{c2}$ (5\,T curves in Fig.\,\ref{fig1}), $\alpha_i/T$ do not show any strong divergence down to 100\,mK that could be related to quantum critical behavior, ruling out the presence of nearby pressure-induced instabilities.

\begin{figure}[t]
\includegraphics[width=3in]{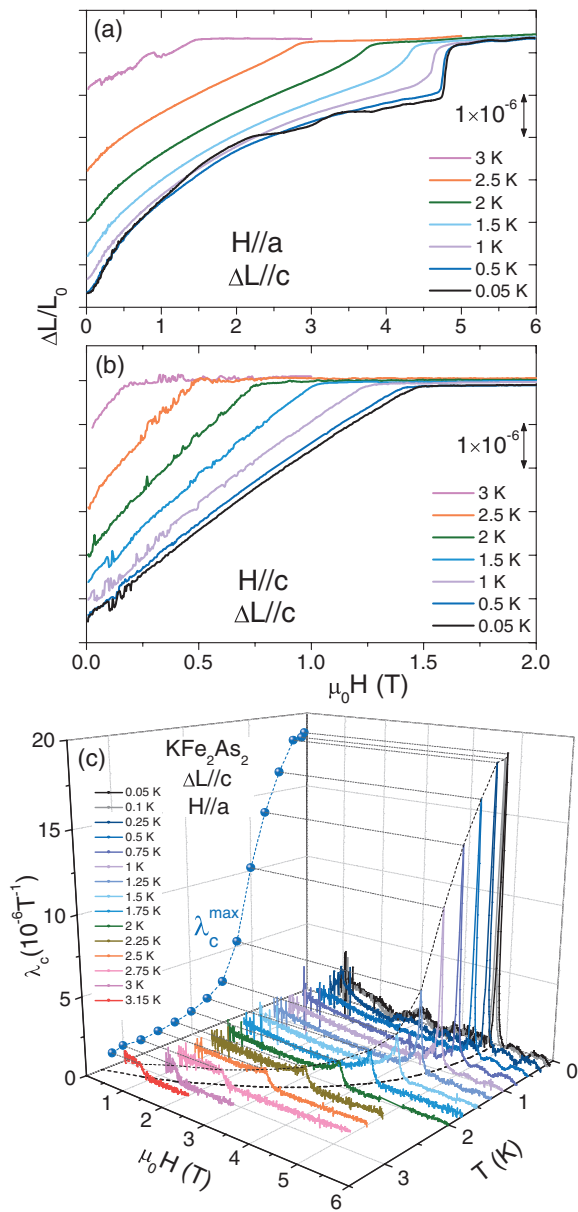}
\caption{(color online). Changes in sample length $\Delta L$\,=\,$L - L_{0}$ measured along the $c$ axis of the crystal versus magnetic field (a) $H$\,$\parallel$\,$a$ and (b) $H$\,$\parallel$\,$c$ at temperatures ranging from 0.05\,K (lowest curve) to 3\,K (uppermost curve) ($L_{0}$\,=\,500\,$\mu$m). (c) Magnetostriction $\lambda_c$ vs $H$\,$\parallel$\,$a$ for 0.05\,K\,$\leq$\,$T$\,$\leq$\,3\,K. $\lambda_c$ maxima are plotted in the projected $\lambda_c$--$T$ plane (blue circles), from which a tricritical temperature \ttt\,$\sim$\,1.5\,K can be extracted.}
\label{fig2}
\end{figure}

\begin{figure}[t]
\includegraphics[width=3.2in]{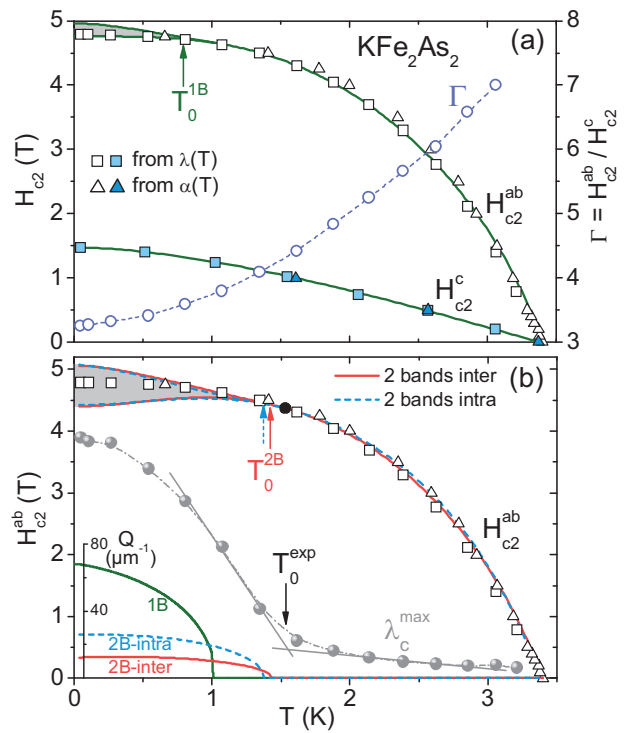}
\caption{(color online). (a) Left axis: $H_{c2}^{ab}$ (open symbols) and $H_{c2}^{c}$ (closed symbols) vs $T$ for \kfa, determined from $\alpha(T)$ (triangles) and $\lambda(T)$ (squares). Solid lines correspond to single-band calculations, with the vertical arrow indicating the position of the corresponding tricritical-point temperature $T_0^{\mathrm{1B}}$, below which the FFLO phase is predicted to form. Right axis: the $H_{c2}$ anisotropy factor $\Gamma$ vs $T$ (circles). (b) Two-band calculations for $H$\,$\parallel$\,$ab$. The limiting cases of dominant interband (intraband) coupling are indicated by solid (dashed) lines. Arrows indicate the position of the tricritical-point temperature from $\lambda_c^{max}$ ($T_0^{exp}$) and from the calculations ($T_0^{\mathrm{2B}}$). $Q$-vector amplitudes vs $T$ obtained from single- and two-band calculations are also displayed. In (a) and (b), the calculated upper lines below \ttt\ represent the onset of the FFLO state, while the lower line corresponds to the onset of the homogeneous phase with $Q$\,=\,0.}
\label{fig3}
\end{figure}

The SO--FO crossover becomes strikingly visible in the magnetostriction data displayed in Fig.\,\ref{fig2}. For $H$\,$\parallel$\,$a$, a discontinuous variation of the sample length develops at $H_{c2}^{ab}$ as the field is swept at low-$T$ [Fig.\,\ref{fig2}(a)]. Clearly, this discontinuity is not present for $H$\,$\parallel$\,$c$ [Fig.\,\ref{fig2}(b)]. The first-order-like length discontinuities observed at low temperatures for $H$\,$\parallel$\,$a$ translate into the very pronounced peaks of the length derivatives $\lambda_c$($H,T$) displayed in Fig.\,\ref{fig2}(c). At 50\,mK, the maximum value of $\lambda_c$($H$) is almost 20 times larger than the transition step at 3\,K. The values of $\lambda_c^{max}$ are plotted in the projected $\lambda_c$--$T$ plane, from which it is possible to define a SO--FO crossover temperature \ttt\,$\sim$\,1.5\,K [see also Fig.\,\ref{fig3}(b)]. Magnetic-field hysteresis is also observed at low temperatures, consistent with the FO character of the transition, and appears to be suppressed above 500\,mK \cite{sm}. We have ruled out, on the basis of a detailed examination of the hysteretic behavior \cite{sm}, the possibility that the FO transition could arise from the onset of the irreversible regime of the vortex lattice, which appears at magnetic fields slightly smaller than $H_{c2}^{ab}$ at 50\,mK and persists even where SO superconducting transitions are observed, for example, at 2.5\,K for $H$\,$\parallel$\,$a$ and 50\,mK for $H$\,$\parallel$\,$c$.

The $H$--$T$ phase diagram derived from our $\alpha_i$ and $\lambda_i$ measurements is presented in Fig.\,\ref{fig3}(a). While $H_{c2}^c$ increases monotonically with decreasing $T$, $H_{c2}^{ab}$ flattens out below 1.5\,K, a sign of strong Pauli-limiting effects. Moreover, $H_{c2}^{ab}(0)$\,=\,4.8\,T, which is much smaller than the clean-limit orbital field 0.73\,$T_c\,dH_{c2}/dT|_{T_c}$$\sim$\,15.4\,T. The crossover to a discontinuous phase transition at \ttt\ can only be observed for $H$\,$\parallel$\,$ab$ (Figs.\,\ref{fig1} and \ref{fig2}), the field direction for which shielding currents are minimal. This suggests that the driving force for Pauli limitation in \kfa\ is the quasi-two-dimensional electronic structure, in contrast to CeCoIn$_5$ where FO transitions appear for both field directions \cite{bianchi03a}. Despite the clear indications of Pauli limitation in $H_{c2}^{ab}$, our data does not show signatures of a possible FFLO phase at high fields such as a double transition or an upturn of $H_{c2}(T)$ towards low $T$ observed in $\kappa$-(BEDT-TTF)$_2$Cu(NCS)$_2$, where BEDT-TTF is bisethylenedithio-tetrathiafulvalene \cite{lortz07a}. We cannot rule out the possibility of a slight misalignment of the magnetic field with the sample inhibiting the formation of the FFLO phase \cite{beyer12a}. Remarkably, $H_{c2}^{ab}$ and $H_{c2}^{c}$ of \kfa\ exhibit a $T$-dependent anisotropy factor $\Gamma$\,=\,$H_{c2}^{ab}/H_{c2}^{c}$ [Fig.\,\ref{fig3}(a)], contrary to the constant anisotropy expected from Ginzburg-Landau theory. This unusual anisotropy has also been reported for LiFeAs and Fe(Se,Te) \cite{braithwaite10a,kurita11a,cho11a,kogan12a}, and has been attributed to Pauli limiting and/or multiband effects.

To interpret theoretically our results, we model \hc\ using first a single-band formalism. We considered the solutions to the linearized Gor'kov equations developed by Werthamer, Helfland and Hohenberg (WHH) for a uniaxial, clean-limit SC, following the approach recently presented by Gurevich \cite{WH1966a,WH1966b,gurevich10a}. This model takes into account orbital and Zeeman pair-breaking effects, as well as the formation of an FFLO state below a tricritical temperature \ttt\ when its modulation wavelength $\lambda_Q$ is shorter than the mfp $\ell$. Apart from the \hc\ curve, the model yields the Fermi velocities and the Pauli susceptibility $\chi_N$\,=\,$(1/2) g^2 \mu_\mathrm{B}^2 N(E_\mathrm{F})$ by obtaining the gyromagnetic factor $g$. It also determines the FFLO phase boundaries below \ttt\ and the modulation vector $Q$\,$\propto$\,$\lambda_Q^{-1}$, although these values should be taken with caution as they are sensitive to details of the electronic band structure and disorder which are not considered in the model. The values of $v_{F}$ were always kept within the range of the values deduced from our QOs (see \cite{sm} for a summary of the parameters used in the calculations).

The calculated single-band \hc\ curves are displayed in Figs.\,\ref{fig3}(a). They adjust well the experiment for both field directions. For $H$\,$\parallel$\,$ab$, the calculations give a Maki parameter $\alpha_\mathrm{M}$\,=\,3.8, consistent with the observation of a FO phase transition. On the other hand, the calculations predict $T_0^{\mathrm{1B}}$\,$\sim$\,1\,K, significantly smaller than $T_0^{exp}$\,$\sim$\,1.5\,K determined from $\lambda_c^{max}$. This is remarkable, as the calculated value should constitute an upper limit: \ttt\ is determined by $\alpha_\mathrm{M}$, $i.\,e.$ the balance between Pauli and orbital pair-breaking. \ttt\ is hardly changed by antiferromagnetic (AFM) fluctuations, nodes in the gap function, or strong-coupling effects \cite{brison97a}. Disorder, on the other hand, suppresses the Pauli pair-breaking effects and reduces \ttt\ \cite{terashima09a}. The discrepancy between a rather high $T_0^{exp}$ and the single-band $\alpha_\mathrm{M}$ is further illustrated by a comparison with other Pauli-limited SCs. The organic SC $\kappa$-(BEDT-TTF)$_2$Cu(NCS)$_2$, for example, has a comparable temperature $t_0$\,=\,$T_0/T_c$, determined by the peak height of $C/T$, but a much higher $\alpha_\mathrm{M}$\,=\,8 \cite{lortz07a}. In a clean-limit SC, the orbital pair-breaking effects are determined by the Fermi velocity of the shielding currents which can be extracted from the slope $H_{c2}^{'}$ at \tc. The deviation of $H_{c2}$(0) from $H_{c2}^{orb}$, on the other hand, represents a measure of Pauli-limiting effects. As \tc, $H_{c2}^{'}$, and $H_{c2}$(0) are the only free parameters in the model, the irreconcilable difference between $T_0^{exp}$ and $T_0^{\mathrm{1B}}$ suggests the impossibility of describing \kfa\ with a single-band model.

The failure of the single-band model in explaining the $H$--$T$ phase diagram of \kfa\ becomes even more evident for $H$\,$\parallel$\,$c$, where calculations result in the unphysical value $g_{c}$\,$\approx$\,0, although values of $g$\,$<$\,2 might be possible in the presence of the Jaccarino-Peter effect \cite{JP62a}. Since AFM fluctuations are indeed present in \kfa\ with the magnetic easy plane perpendicular to the $c$ axis, a reduced $g$-factor should be visible for $H$\,$\parallel$\,$ab$ and not for $H$\,$\parallel$\,$c$. Furthermore, the extreme magnetic anisotropy indicated by $g_{ab} / g_{c}$\,$\rightarrow$\,$\infty$ as a result of a single-band model clearly contradicts magnetization and Knight-shift measurements which unambiguously reveal a nearly isotropic susceptibility, with $\chi_{ab} / \chi_{c}$\,$\approx$\,1.2--1.5 \cite{zhang10a,hirano12a,hardy13a}.

If more than one band contributed to the FS, the slope $H_{c2}^{'}$ would be proportional to a superposition of Fermi velocities, $H_{c2}^{'}$\,$\propto$\,$(\sum_n c_n v_{F,n})^{-1}$. The coefficients $c_n$ are functions of the superconducting coupling constants. Since bands differing in shape and dimensionality could essentially yield very different values of $v_{F,n}$, it is very well conceivable that one band could be Pauli limited while the others remained orbitally limited. In this case, a SO--FO crossover would occur below a high value of \ttt\ even for a relatively small slope $H_{c2}^{'}$. The FFLO state, on the other hand, could be damped by the bands with dominating orbital pair-breaking. The electronic structure of \kfa\ inferred from our QOs and recent ARPES measurements does indeed reveal bands of different characteristics \cite{terashima10a,okazaki12a}. The five bands that cross the Fermi energy and hence contribute to the FS, however, cannot be considered in the model due to the complexity of the calculations. In order to capture the basic physical description, we restrict the calculations to a two-band model, for which four coupling constants enter as additional parameters. We therefore adjust the data within two extreme scenarios: one dominated by interband coupling, and the other by intraband coupling, as proposed for Fe-based SCs and for MgB$_2$, respectively. The latter scenario is supported by our thermal-expansion measurements which show similarities with this material.

The results of the two-band calculations are presented in Fig.\,\ref{fig3}(b), showing that this model moves $T_0^{\mathrm{2B}}$ to higher temperatures compatible with the experiment, while keeping $H_{c2}^{'}$\,$\sim$\,--\,6\,T/K. With these parameters, $H_{c2}^{ab}(T)$ is practically independent of the coupling constants. The particular multiband topology of the Fermi surface of \kfa\ results in a higher crossover temperature. The higher $T_0^{\mathrm{2B}}$ leads to a more extended stability range of the FFLO state compared to the single-band calculations. The band which is less affected by Pauli limiting inhibits, however, the formation of an FFLO state (smaller $Q$) in the multi-band case, resulting in a larger value of $\lambda_Q$ which exceeds $\ell_{ab}$ \cite{sm}. The effect of suppression of the FFLO phase by non-Pauli-limited bands is expected to be stronger if all five bands involved in the electronic structure of \kfa\ are considered in the model.

Multiband superconductivity is ubiquitous in Fe-based superconductors. In \bkfa, increasing the K content lowers the dimensionality of the electronic structure and gives rise to strong correlations. These conditions favor Pauli pair-breaking effects in \kfa, where we found compelling evidence for Pauli-limited multiband superconductivity. In more general terms, our experiments have shown the complex interplay of pair breaking and multiband effects, which have to be taken into account in models of multiband superconductivity in iron-based superconductors \cite{mizushima13a,ptok13a}.

The authors thank A. Gurevich, J. Wosnitza, G. Zwicknagl, C. Meingast, P. Burger, F. Hardy, R. Hott, R. Eder and J. Schmalian for stimulating discussions, and R. Sch\"{a}fer and S. Zaum for help with the experiments. This work has been partially supported by the DFG through SPP1458.

\end{document}


\newcommand{\kfa}{KFe$_2$As$_2$}
\newcommand{\bkfa}{Ba$_{1-x}$K$_x$Fe$_2$As$_2$}
\newcommand{\tc}{$T_c$}

\title{Pauli-Limited Multiband Superconductivity in \kfa}

\author{D. A. Zocco}
\author{K. Grube}
\author{F. Eilers}
\author{T. Wolf}
\affiliation{Institute of Solid State Physics (IFP), Karlsruhe Institute of Technology, D-76021 Karlsruhe, Germany.}
\author{H. v.\,L\"{o}hneysen}
\affiliation{Institute of Solid State Physics (IFP), Karlsruhe Institute of Technology, D-76021 Karlsruhe, Germany.}
\affiliation{Physikalisches Institut, Karlsruhe Institute of Technology, D-76031 Karlsruhe, Germany}

\maketitle

\section{I. SUPPLEMENTAL MATERIAL}

\subsection{A. SAMPLE PREPARATION}

Single crystals of \kfa\ were grown from self flux. With FeAs flux, only samples of minor quality and small size were obtained. Fluxes rich in As and K yield high quality single-crystals of large size ($\sim$\,5\,$\times$\,5\,mm$^2$). The appropriate amounts of K, As, and pre-reacted FeAs or FeAs$_2$ were combined in an Al$_2$O$_3$ crucible and sealed in a steel cylinder under 1\,bar of argon. Following the heating of the cylinder to 920\,--\,1000$^{\circ}$C, the crystal growth was started by slowly cooling the furnace to 730\,--\,850$^{\circ}$C at rates of 0.3\,--\,0.49$^{\circ}$C/hour. Once the lower temperature was reached, the cylinder was tilted to separate the crystals from the remaining liquid flux, followed by a slow cool-down to room temperature. After opening the containers, it was possible to collect the free-standing single crystals with tweezers, while an etching treatment with ethanol was necessary to separate the remaining crystals trapped in the solid flux. Samples with typical residual-resistivity ratios RRR\,=\,$\rho$(300\,K)/$\rho$(4\,K)\,$\sim$\,1000 were obtained \cite{hardy13a}.

\subsection{B. QUANTUM OSCILLATIONS IN MAGNETOSTRICTION}

\begin{figure}[h]
\begin{center}
\includegraphics[width=0.9\textwidth]{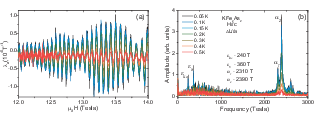}
\caption{(a) High-field magnetostriction $\lambda_a$ of \kfa\ for magnetic fields $H$ applied along the $c$--axis at several temperatures. (b) Fourier-transform spectra of the quantum oscillations displayed in (a).}
\label{fig1s}
\end{center}
\end{figure}

Fig.\,\ref{fig1s} displays the quantum oscillations (QOs) observed in magnetostriction measurements of the sample length along the $a$--axis of the sample and for applied magnetic fields $H$\,$\parallel$\,$c$. Peaks are labeled following previous de Haas--van Alphen (dHvA) measurements \cite{terashima10a}. Details of the analysis of the QOs data (effective masses, mean free paths, etc.) will be presented in a separate manuscript.

\subsection{C. HYSTERESIS IN MAGNETOSTRICTION}

\begin{figure}[h]
\begin{center}
\includegraphics[width=0.6\textwidth]{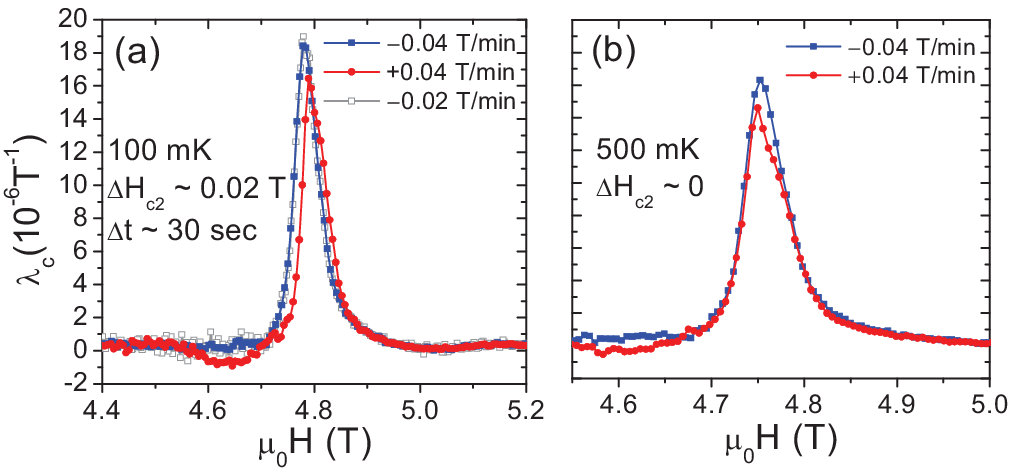}
\caption{Magnetostriction $\lambda_c$ vs.\,field at (a) 100\,mK and (b) 500\,mK.}
\label{fig2s}
\end{center}
\end{figure}

The data displayed in Fig.\,\ref{fig2s}(a) shows a clear shift of the rising and trailing edges between the 100\,mK magnetostriction curves obtained upon increasing and decreasing magnetic field ($H$\,$\parallel$\,$a$), which we attribute to hysteresis effects associated with the first-order phase transition. This hysteresis appears to be suppressed above 500\,mK [Fig.\,\ref{fig2s}(b)] except for small changes in the height of the maxima. $H_{c2\uparrow}$ obtained upon increasing $H$ at a rate $\mu_0 \dot{H}$\,$=$\,0.04\,T/min is shifted with respect to $H_{c2\downarrow}$ measured while decreasing $H$ at $\mu_0 \dot{H}$\,$=$\,$-$0.04\,T/min. The shift $\mu_0 \Delta H_{c2}$\,$=$\,$\mu_0\left| H_{c2\uparrow}-H_{c2\downarrow}\right|$\,$\sim$\,0.02\,T corresponds to a time shift $\Delta t$\,$\sim$\,30\,seconds for $\mu_0 \dot{H}$\,$=$\,0.04\,T/min, which cannot be attributed to a measurement delay since it is much higher than the delay of our apparatus, reflected in the almost identical curves measured with different rates, $\mu_0\dot{H}$\,$=$\,$-$0.04\,T/min and $\mu_0 \dot{H}$\,$=$\,$-$0.02\,T/min.

\begin{figure}[h]
\begin{center}
\includegraphics[width=1\textwidth]{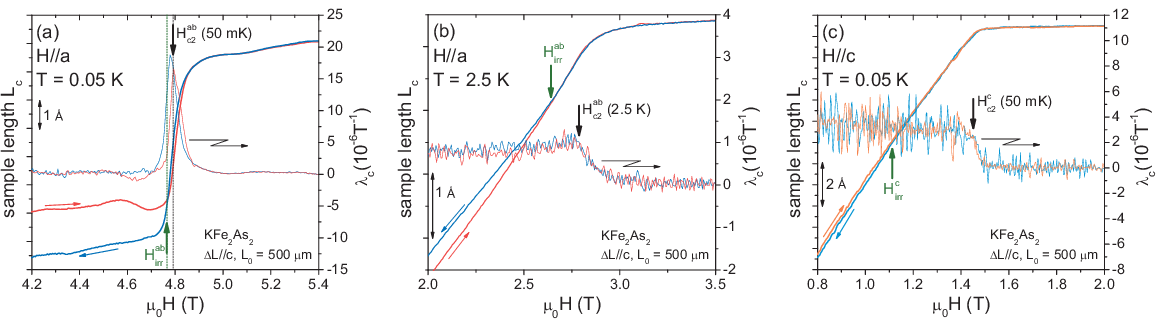}
\caption{$c$-axis sample length $L_c$ (left axis) and magnetostriction $\lambda_c$ (right axis) versus applied magnetic field $H$. (a) $H$\,$\parallel$\,$a$, $T$\,=\,50\,mK; (b) $H$\,$\parallel$\,$a$, $T$\,=\,2.5\,K; (c) $H$\,$\parallel$\,$c$, $T$\,=\,50\,mK. The green arrows indicate the value of magnetic field below which the measurements taken during decreasing (blue) and increasing (red) magnetic field separate.}
\label{fig3s}
\end{center}
\end{figure}

The hysteresis is also observed directly in the length change, as shown in more detail in Fig.\,\ref{fig3s}(a) for $H$\,$\parallel$\,$a$ and $T$\,=\,50\,mK. On the other hand, the curves measured in decreasing (blue) and increasing (red) magnetic field separate notoriously right below the superconducting transition at $H_{c2}^{ab}$(50\,mK) (green arrow), which might indicate the entrance into the irreversible vortex state. The splitting of the two curves persists in the measurements taken at $T$\,=\,2.5\,K for $H$\,$\parallel$\,$a$ [Fig.\,\ref{fig3s}(b)] and at $T$\,=\,50\,mK for $H$\,$\parallel$\,$c$ [Fig.\,\ref{fig3s}(c)], where the superconducting transition is always of second order.

\subsection{D. UPPER CRITICAL FIELD CALCULATIONS}

The calculations of $H_{c2}(T)$ were performed using the model developed by Werthamer, Helfland and Hohenberg (WHH) for a uniaxial, clean-limit SC \cite{WH1966a,WH1966b}, following the approach recently presented by A. Gurevich \cite{gurevich10a}. For the single-band (1B) calculations the equation
\begin{equation}\label{gure1}
\mathrm{ln}\,t + U(t,b,q)=0
\end{equation}
has to be solved, which involves the calculation of $U(t,b,q)$
\begin{equation}\label{gure2}
U(t,b,q) = 2\mathrm{e}^{q^2} \mathrm{Re} \sum_{n=0}^{\infty}\int_q^{\infty} du\, \mathrm{e}^{-u^2}
\left\{ \frac{u}{n+1/2}-\frac{t}{\sqrt{b}} \mathrm{tan}^{-1} \left[ \frac{u \sqrt{b}}{t ( n+1/2) + \mathrm{i}\alpha_\mathrm{G} b}\right] \right\}
\end{equation}
where $t$\,=\,$T/T_c$, and $b$, $q$ and $\alpha_\mathrm{G}$ are the reduced upper critical field, reduced magnitude of the $Q$-vector corresponding to the FFLO phase, and reduced Maki parameter, respectively. In this model, an FFLO phase appears if $\alpha_\mathrm{G}$\,$\geq$\,1, and it is related to the actual Maki parameter via $\alpha_\mathrm{G}$\,$\approx$\,$\alpha_\mathrm{M}$\,/\,1.845. The band anisotropy can be described in terms of the ratio of the effective masses or the ratio of the Fermi velocities:
\begin{equation}\label{gure3}
\epsilon = \frac{m_{\perp}}{m_{\parallel}} = \left( \frac{v_{\parallel}}{v_{\perp}} \right)^2
\end{equation}
where $\parallel$ and $\perp$ denote the directions parallel and perpendicular to the direction of the applied magnetic field:
\begin{equation}\label{gure4}
for \quad H\parallel c\,: \quad \qquad v_{\parallel}=v_{c},\quad v_{\perp}=v_{ab} \qquad \rightarrow \qquad \epsilon_{\parallel c} = \left( \frac{v_{c}}{v_{ab}} \right)^2
\end{equation}
\begin{equation}\label{gure5}
for \quad H\parallel ab\,: \quad \qquad v_{\parallel}=v_{ab},\quad v_{\perp}=\sqrt{v_{ab}v_{c}} \qquad \rightarrow \qquad \epsilon_{\parallel ab} = \frac{v_{ab}}{v_{c}}
\end{equation}

The shielding currents are determined by the component of the velocity perpendicular to the field, $i.\,e.$, $\frac{dH_{c2}}{dT}|_{T_c}$\,$\propto$\,$\frac{1}{v_{\perp}}$.

For a two-band (2B) calculation, we need to solve the following equation:
\begin{equation}\label{gure6}
a_1 (\mathrm{ln}\, t + U_1)+a_2 (\mathrm{ln}\, t + U_2)+(\mathrm{ln}\, t + U_1)(\mathrm{ln}\, t + U_2)=0
\end{equation}
where $U_1$ corresponds to Eq.~\ref{gure2} and $U_2$ is defined as
\begin{equation}\label{gure7}
U_2(t,b,q) = 2\mathrm{e}^{q^2 s}\, \mathrm{Re} \sum_{n=0}^{\infty}\int_{q\sqrt{s}}^{\infty} du\, \mathrm{e}^{-u^2}
\left\{ \frac{u}{n+1/2}-\frac{t}{\sqrt{b\eta}} \mathrm{tan}^{-1} \left[ \frac{u \sqrt{b\eta}}{t ( n+1/2) + \mathrm{i}\alpha_\mathrm{G} b}\right] \right\}
\end{equation}
with the interband parameters $s$ and $\eta$ defined as
\begin{equation}\label{gure8}
s = \frac{\epsilon_{2}}{\epsilon_{1}}, \qquad \eta = \left( \frac{v_{ab,2}}{v_{ab,1}} \right)^2.
\end{equation}

The parameters $a_1$ and $a_2$ in Eq.~\ref{gure6} are defined as:
\begin{equation}\label{gure9}
a_1 = \frac{\lambda_0 + \lambda_-}{2w},\quad a_2 = \frac{\lambda_0 - \lambda_-}{2w}
\end{equation}
\begin{equation}\label{gure10}
\lambda_- = \lambda_{11} - \lambda_{22},\quad \lambda_0 = \sqrt{\lambda_-^2 + 4\lambda_{12}\lambda_{21}},\quad w = \lambda_{11}\lambda_{22}-\lambda_{12}\lambda_{21}.
\end{equation}
and determine the level of intraband and interband coupling of the calculations:
\begin{itemize}
  \item $\lambda_{11}\lambda_{22}$\,$<<$\,$\lambda_{12}\lambda_{21}$ \quad $\rightarrow$ \quad interband coupling limit (2B-inter)
  \item $\lambda_{11}\lambda_{22}$\,$>>$\,$\lambda_{12}\lambda_{21}$ \quad $\rightarrow$ \quad intraband coupling limit (2B-intra)
\end{itemize}

The following tables summarize the parameters used in the single- and two-band calculations, where $g$ corresponds to the gyromagnetic factor, and $\lambda_Q^{min}$  is proportional to the inverse of the maximum value of $Q$ displayed in Fig.\,\ref{fig4s}(a).

{\renewcommand{\arraystretch}{1.2}
\begin{table}[h]
\caption{\label{table1}
Single-band calculation parameters.}
\centering
\begin{tabular}{c || c | c | c | c | c | c | c | c | c | c | c}
\cline{1-12}
$H$-field & {\multirow{2}{*}{~$T_c$ (K)~}} & {\multirow{2}{*}{~~$g$~~}}  & {\multirow{2}{*}{ $v_{ab}$ (m/s) }}  & {\multirow{2}{*}{ $v_{c}$ (m/s) }}  & {\multirow{2}{*}{$\frac{dH_{c2}}{dT}|_{T_c}$ (T/K)}} & {\multirow{2}{*}{$H_{c2}^{orb}$ (T)}} & {\multirow{2}{*}{$H_{c2}^{\mathrm{P}}$ (T)}} & {\multirow{2}{*}{~~$\alpha_\mathrm{M}$~~}} & {\multirow{2}{*}{$T_0^{\mathrm{1B}}$ (K)}} & {\multirow{2}{*}{$t_0$\,=\,$\frac{T_0^{\mathrm{1B}}}{T_c}$}} & {\multirow{2}{*}{$\lambda_Q^{min}$ (nm)}} \\
direction &      &      &       &      &        &       &         &        &       &     \\
\cline{1-12}
   $ab$   &{\multirow{2}{*}{3.4}}& 2.24 &  42500  & 4100 &  -6.2  & 15.33  &  5.65  & 3.84 & 1.01 & 0.295 & 52 \\
   $c$    &                      &  0   &  42500  & 4100 &  -0.6  & ~1.48  &$\infty$&  0   &  --  &   --  & -- \\
\cline{1-12}
\end{tabular}
\end{table}

\vspace{10mm}

{\renewcommand{\arraystretch}{1.2}
\begin{table}[h!]
\caption{\label{table2}
Two-band calculation parameters, for strong interband coupling scenario.}
\centering
\begin{tabular}{c || c | c | c | c || c | c | c | c | c | c}
\cline{1-11}
$H$-field &  \multicolumn{4}{c||}{Coupling parameters}  & {\multirow{2}{*}{~$T_c$ (K)~}} &  {\multirow{2}{*}{~~~$g$~~~}} & {\multirow{2}{*}{~$\frac{dH_{c2}}{dT}\bigg|_{T_c}$ (T/K)~}} &  {\multirow{2}{*}{~$T_0^{\mathrm{2B}}$ (K)~}} & {\multirow{2}{*}{$t_0$\,=\,$\frac{T_0^{\mathrm{2B}}}{T_c}$}} & {\multirow{2}{*}{$\lambda_Q^{min}$ (nm)}}\\
\cline{2-5}
direction & ~~~$\lambda_{11}$~~~ & ~~~$\lambda_{22}$~~~ & ~~~$\lambda_{12}$~~~ & ~~~$\lambda_{21}$~~~ &      &      &      &       &       &     \\
\cline{1-11}
   $ab$   &{\multirow{2}{*}{0}}&{\multirow{2}{*}{0}}&{\multirow{2}{*}{0.5}}&{\multirow{2}{*}{0.5}}&{\multirow{2}{*}{3.4}}& 2.23 & -5.99 & 1.43 & 0.42 & 495 \\
   $c$    &     &     &    &    &    &   2  & -0.332&  --   & --   & -- \\
\cline{1-11}
\end{tabular}
\newline
\newline
\begin{tabular}{c || c | c | c | c | c || c | c | c | c | c }
\cline{1-11}
$H$-field &  \multicolumn{5}{c||}{Band 1}  &  \multicolumn{5}{c}{Band 2}    \\
\cline{2-11}
direction & $v_{ab}$ (m/s) & $v_{c}$ (m/s) &  $H_{c2}^{orb}$ (T) & $H_{c2}^{\mathrm{P}}$ (T) & ~~$\alpha_\mathrm{M}$~~ & $v_{ab}$ (m/s) & $v_{c}$ (m/s) &  $H_{c2}^{orb}$ (T) & $H_{c2}^{\mathrm{P}}$ (T) & ~~$\alpha_\mathrm{M}$~~ \\
\cline{1-11}
   $ab$   &  28000 & 11700 & ~8.15 & 5.67 & ~2.03 & 57000 & ~574 & 81.50 & 5.67 & 20.35 \\

   $c$    &  28000 & 11700 & ~3.40 & 6.31 & ~0.76 & 57000 & ~574 & ~0.82 & 6.31 & ~0.18 \\
\cline{1-11}
\end{tabular}
\end{table}

\vspace{10mm}

\begin{table}[h!]
\caption{\label{table3}Two-band calculation parameters, for strong intraband coupling scenario.}
\centering
\begin{tabular}{c || c | c | c | c || c | c | c | c | c | c}
\cline{1-11}
$H$-field &  \multicolumn{4}{c||}{Coupling parameters}  & {\multirow{2}{*}{~$T_c$ (K)~}} &  {\multirow{2}{*}{~~~$g$~~~}} & {\multirow{2}{*}{~$\frac{dH_{c2}}{dT}\bigg|_{T_c}$ (T/K)~}} &  {\multirow{2}{*}{~$T_0^{\mathrm{2B}}$ (K)~}} & {\multirow{2}{*}{$t_0$\,=\,$\frac{T_0^{\mathrm{2B}}}{T_c}$}} & {\multirow{2}{*}{$\lambda_Q^{min}$ (nm)}}\\
\cline{2-5}
direction & ~~~$\lambda_{11}$~~~ & ~~~$\lambda_{22}$~~~ & ~~~$\lambda_{12}$~~~ & ~~~$\lambda_{21}$~~~ &      &      &      &       &       &     \\
\cline{1-11}
   $ab$   &{\multirow{2}{*}{0.2}}&{\multirow{2}{*}{0.8}}&{\multirow{2}{*}{0.2}}&{\multirow{2}{*}{0.2}}&{\multirow{2}{*}{3.4}}& 2.33 & -4.48 & 1.38 & 0.41 & 244 \\
   $c$    &     &     &    &    &    &   2  & -0.533&  --   &  --  & -- \\
\cline{1-11}
\end{tabular}
\newline
\vspace*{2mm}
\newline
\begin{tabular}{c || c | c | c | c | c || c | c | c | c | c }
\cline{1-11}
$H$-field &  \multicolumn{5}{c||}{Band 1}  &  \multicolumn{5}{c}{Band 2}    \\
\cline{2-11}
direction & $v_{ab}$ (m/s) & $v_{c}$ (m/s) &  $H_{c2}^{orb}$ (T) & $H_{c2}^{\mathrm{P}}$ (T) & ~~$\alpha_\mathrm{M}$~~ & $v_{ab}$ (m/s)  & $v_{c}$ (m/s) &  $H_{c2}^{orb}$ (T) & $H_{c2}^{\mathrm{P}}$ (T) & ~~$\alpha_\mathrm{M}$~~ \\
\cline{1-11}
   $ab$   & 18000 & 115000 & ~1.29 & 5.42 & ~0.34 & 45000 & 1644 & 36.1~~ & 5.42 & ~9.44 \\
   $c$    & 18000 & 115000 & ~8.24 & 6.31 & ~1.85 & 45000 & 1644 & ~1.32 & 6.31 & ~0.30 \\
\cline{1-11}
\end{tabular}
\end{table}

\begin{figure}[h!]
\begin{center}
\includegraphics[width=0.8\textwidth]{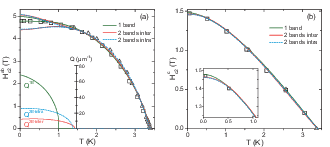}
\caption{One-band (1B) and two-band (2B) calculations for (a) $H$\,$\parallel$\,$ab$ and (b) $H$\,$\parallel$\,$c$. }
\label{fig4s}
\end{center}
\end{figure}